\documentclass[12pt]{article}
\usepackage{a4wide}
\usepackage{latexsym}
\usepackage{amsfonts}

\def\bq{\begin{eqnarray}}
\def\eq{\end{eqnarray}}
\def\l{\langle}
\def\r{\rangle} 
\def\eps{\varepsilon}

\newlength{\dinwidth} \newlength{\dinmargin}
\begin{document}

\thispagestyle{empty}

\begin{flushright}
  UPRF-2001-14
\end{flushright}

\vspace{1.5cm}

\begin{center}
 {\Large\bf A general algorithm to generate
unweighted events for next-to-leading order calculations in electron-positron annihilation\\}
  \vspace{1cm}
  {\sc Stefan Weinzierl}\\
  \vspace{1cm}
  {\it Dipartimento di Fisica, Universit\`a di Parma,\\
       INFN Gruppo Collegato di Parma, 43100 Parma, Italy} \\
\end{center}

\vspace{2cm}

% abstract ---------------------------------------
\begin{abstract}\noindent
{ 
Given a next-to-leading order calculation, we show how to set up a computer
program, which generates a sequence of unweighted momentum configurations, each
configuration containing either $n$ or $n+1$ four-vectors, such that for any infrared
safe observable the average over these configurations coincides with the NLO
calculation up to errors of order $y_{res}$.
The core of the algorithm is a method to combine real emission and virtual corrections
on a point-by-point basis in hard phase space.
The algorithm can be implemented on top of existing NLO calculations.
}
\end{abstract}

\vspace*{\fill}

% main text ------------------------------------
\newpage

\reversemarginpar

\section{Introduction}

As high energy experiments move towards a new generation of colliders, accurate
simulation of collision events becomes increasingly important.
Among the main tools used today are next-to-leading order (NLO) calculations 
and event generators.
Up to now these two tools have a complementary regions of applicability.
A NLO-calculation for an $n$-jet observable calculates this observable as a sum
of two terms, one term living on the $n$-parton phase space, the other on the
$(n+1)$-parton phase space. These two contributions correspond roughly to the
virtual corrections and the real emissions, where infrared divergences have been
taken care of by a method like phase space slicing or the subtraction method.
Usually the two contributions to an observable are only combined in the end.
As a side-effect, one deals at intermediate steps with variable and/or
negative weights.\\
\\
Event generators like HERWIG \cite{Corcella:2000bw} and PYTHIA \cite{Sjostrand:1994yb} on the other hand produce unweighted
events through a three-stage process: From an initial hard process parton showers
are generated. A hadronization model converts at a low scale the partons into
hadrons.
The hard process is usually based on Born matrix elements.\\
\\
The two approaches have complementary strenght: Parton showers
describe quite reasonably multiple collinear emission, whereas a NLO-calculation
is by far more accurate when $n$ partons are well separated in phase space.\\
\\
It is a topic of current interest to merge the benefits of the two approaches
into an ``event generator accurate at NLO''.
Friberg and Sj\"ostrand have suggested a possible road in \cite{Friberg:1999fh}.
Similar approaches have been considered
%Their suggestions have been taken up for specific processes 
by Collins \cite{Collins:2000qd},
P\"otter \cite{Potter:2000an} and Dobbs and Lefebvre \cite{Dobbs:2000bx} and applied to deep-inelastic scattering
(\cite{Collins:2000gd}, \cite{Potter:2001ej}) and the production of electroweak gauge
bosons in hadron-hadron collisions (\cite{Dobbs:2001gb}, \cite{Chen:2001ci}).
Mrenna has considered the case how to combine resummed calculations with parton showers
\cite{Mrenna:1999mq}.
Despite this progress, there is still room for improvement: 
The algorithm of Collins may generate events with negative weight.
The algorithm of P\"otter uses phase space slicing and introduces a unresolved region, such
that the sum of the Born term, 
the virtual corrections and the real emission part is exactly zero \cite{Baer}.
It is not guaranteed that this region is always small enough such that the
approximations of phase space slicing are still valid. \\
\\
In combining parton showers with NLO-calculations two problems can immediately be identified:
Negative weights and double counting.
In this paper we would like to address the first problem.
As already stated before, a specific momentum configuration within a NLO
calculation might come with a negative weight.
There is nothing wrong with this fact by itself and one might think
of carrying this weight along the further simulation of the event.
However, there are some concerns raised against this approach:
\begin{description}
\item[a.)] Some hadronization models have problems with negative weights \cite{Friberg:1999fh}.
\item[b.)] Experimentalists are unwilling to accept negative weights. Event reconstruction
and detector simulation are quite expensive in terms of computer time and an event
with a negative weight would ``annihilate'' parts of the CPU-time already spent.
\end{description}
Therefore it is desirable to avoid negative weights from the very beginning.
In view of point b.) this is more a question of efficiency than of principle.\\
\\
In this paper we consider an ``NLO partonic event generator''.
Given a next-to-leading order calculation, we show how to set up a computer
program, which generates a sequence of unweighted momentum configurations, each
configuration containing either $n$ or $n+1$ four-vectors, such that for any infrared
safe observable the average over these configurations coincides with the NLO
calculation up to errors of a resolution variable $y_{res}$.
The basic idea is simple: Consider a $(n+1)$-parton configuration. If all
partons are separated by at least $y_{res}$, we return this $(n+1)$-parton
configuration.
If on the other hand a parton is not resolved at a scale $y_{res}$, we average
over the real emissions up to the scale $y_{res}$, combine the result with 
the virtual corrections
and return a $n$-parton configuration.
In this paper we show that this can be done in a general and effective way:
Our procedure does not depend on the specific hard process and the integration over
the real emission part involves only a three-dimensional integration.
These two points are important:
Today many NLO-calculations already exist and we would like to be able to use
them without major modifications.
Our algorithm can take as an input subroutines, which correspond to the
matrix elements for real emission and virtual corrections without modifications.
Only the subtraction terms within the dipole formalism have to be implemented.
Secondly, if a parton is not resolved we have to integrate out this contribution.
An important ingredient is a factorization of the phase space into a 
``hard'' $n$-parton configuration and the insertion of a ``soft'' particle into this
``hard'' configuration.
This allows us to integrate out the unresolved region on a point-by-point bases
in hard phase space.  
We discuss explicitly the case of massless QCD processes in electron-positron
annihilation, although the extension to initial-state partons and/or massive partons
is in principle feasible.\\
\\
In this paper we do not consider how to attach parton showers to the events nor do we address
the problem of double counting. This is a separate problem and addressed for example in
\cite{Friberg:1999fh} and \cite{Collins:2000qd}.\\
\\
This paper is organized as follows:
The next section gives a more detailed outline of the algorithm. Section 3 reviews
the structure of infrared divergences and the dipole formalism.
Section 4 deals with basic Monte Carlo techniques and the generation of phase space.
In section 5 we present an algorithm to generate unweighted events according to leading-order
matrix elements. In section 6 this algorithm is generalized to unweighted events
according to next-to-leading order matrix elements.
Finally, section 7 contains the conclusions.

\section{Preliminaries}

We assume that we have at our disposal a NLO-calculation, consisting of
the virtual corrections, symbolically denoted by
\bq
d\sigma^V & = & \l {\cal M}_{Born}^{(n)} | {\cal M}_{1-loop}^{(n)} \r
               +\l {\cal M}_{1-loop}^{(n)} | {\cal M}_{Born}^{(n)} \r,
\eq
the Born amplitude for the emission of an additional parton
\bq
d\sigma^R & = & \l {\cal M}_{Born}^{(n+1)} | {\cal M}_{Born}^{(n+1)} \r,
\eq
as well as colour-ordered partial amplitudes for the Born process.
The partial amplitudes are needed for the dipole formalism.
We denote the Born matrix element by
\bq
d\sigma^B & = & \l {\cal M}_{Born}^{(n)} | {\cal M}_{Born}^{(n)} \r.
\eq
We distinguish three scales:
\begin{itemize}
\item The scale $y_{cut}$: This scale enters directly the definition of a physical
observable. An example would be the measurement of the 3-jet cross section in electron-positron
annihilation with the Durham algorithm at $y_{cut}=0.01$.
Results of physical observables may depend on the scale $y_{cut}$.
\item The scale $y_{res}$: The resolution scale $y_{res}$ gives the lower bound to which 
we might reasonably push $y_{cut}$. 
On the experimental side $y_{res}$ is set by the finite
resolution of the detector. 
We are here concerned with fixed-order
perturbation theory. 
Consider first the situation where we would like to generate a sequence
of $n$ four-vectors according to the leading-order matrix element.
This matrix element is divergent when one parton becomes either soft or collinear.
We would therefore generate predominately events where one parton is soft or collinear.
For any infrared-safe $n$-jet observable these events will be rejected since
they do not pass the jet-defining cuts. This is inefficient.
We therefore introduce a resolution function $\Theta_{y_{res}}$ and generate only events
which are at least resolved at the scale $y_{res}$.
Let us now consider next-to-leading order calculations.
With multiple scales in the problem, this will inevitably lead to
logarithm of ratios of these scales. If the various scales are not of the same magnitude,
these logarithm can become large and spoil the validity of a fixed-order perturbative
expansion.
Consider the following situation in the real emission part of an NLO-calculation:
Assume that we have $n$ partons well separated in phase space with an additional parton close to one of the
first $n$ partons. 
We would like to replace this $(n+1)$-parton configuration 
with an $n$-parton configuration, where the unresolved parton has been integrated
out. 
Integrating out the unresolved parton in the real emission part up to a resolution
scale $y_{res}$ will yield a positive, but divergent contribution. The divergent
part is compensated by the virtual corrections, and the sum of the real emission part
and the virtual corrections is finite, but not necesarrily positive.
The sign will depend on the scale $y_{res}$. If $y_{res}$ is chosen too small, we do not
include enough contributions from the real emission amplitude and the combined result will
be negative. This is unacceptable for a probabilistic interpretation.
We therefore have too chose $y_{res}$ large enough that for a given ``hard''
$n$-parton configuration we can integrate out an additional ``soft'' parton, which
is within $y_{res}$ of one of the ``hard'' partons and the combination of the obtained
real emission part with the virtual corrections gives a positive contribution.
\item The scale $y_{min}$: This is an internal technical parameter 
we introduce for integrating out efficiently unresolved contributions. 
In some place we will neglect contributions of order $y_{min}$, but since
$y_{min} << y_{res}$ the introduced systematic error is negligible against
the approximations made in the previous step.
\end{itemize}
Of course the technical problem is to integrate out efficiently the unresolved region
for a given ``hard'' $n$-parton
configuration.
We show that this involves only a three-dimensional integration.
We take special care that this can be done efficiently:
Our algorithm is based on the subtraction method. This is more efficient than the phase space
slicing approach, since we don't have to reproduce numerically a logarithm of $y_{min}$, 
which cancels in the end.
Furthermore we use a specific remapping of the unresolved phase space, which tends to 
flatten the integrand.
\\
\\
Our method relies on the following ingredients:
\begin{itemize}
\item The structure of the infrared divergences is universal. We may therefore add and 
subtract an appropriately chosen term, such that the integral over the real emission part
in the unresolved region and the virtual corrections become finite.
Explicitly we are using the dipole formalism of Catani and Seymour \cite{Catani:1997vz} together
with slight modifications given by Nagy and Tr\'ocs\'anyi \cite{Nagy:1998bb}.
\item An algorithm to generate the ``hard'' $n$-parton phase space from a set of random
numbers distributed in $[0,1]$.
There are several algorithms on the market and we use the RAMBO algorithm by Kleiss, Stirling and Ellis \cite{Kleiss:1986gy} for this purpose.
\item An algorithm to insert an additional ``soft'' particle into a given ``hard'' configuration. This algorithm is originally due to Kosower and has been implemented into the
NLO 4-jet program MERCUTIO \cite{Weinzierl:1999yf}.
\item An algorithm to generate a sequence of events according to a multi-dimensional
probability density. We use the Metropolis algorithm for that \cite{Metropolis}.
\end{itemize}
In the next two sections we will briefly review these tools.

\section{Review of the structure of infrared divergences}

In this section we briefly review the dipole formalism and the factorization properties
of QCD amplitudes in the soft and collinear limit.
We follow the notation of Catani and Seymour \cite{Catani:1997vz}.\\
\\
The NLO cross section is given by
\begin{eqnarray}
\sigma^{NLO} & = & \int\limits_n d\sigma^B + \int\limits_n d\sigma^V 
    + \int\limits_{n+1} d\sigma^R.
\end{eqnarray}
$d\sigma^B$ represents the LO-contribution, $d\sigma^V$ the virtual corrections and
$d\sigma^R$ the real emission part. Taken separately, both $d\sigma^V$ and $d\sigma^R$
are divergent, only their sum is finite.
Within the subtraction method, one adds and subtracts a suitable chosen term.
The NLO cross section is therefore rewritten as
\begin{eqnarray}
\sigma^{NLO} & = & \int\limits_n d\sigma^B + \int\limits_n \left( d\sigma^V + \int\limits_1 
d\sigma^A \right) + \int\limits_{n+1} \left( d\sigma^R - d\sigma^A \right).
\end{eqnarray}
The approximation $d\sigma^A$ has to fulfill the following requirements:
\begin{itemize}
\item $d\sigma^A$ must be a proper approximation of $d\sigma^R$ such as to have the same pointwise singular behaviour
(in $D$ dimensions) as $d\sigma^R$ itself. Thus, $d\sigma^A$ acts as a local counterterm for $d\sigma^R$ and one
can safely perform the limit $\varepsilon \rightarrow 0$. This defines a cross-section contribution
\begin{eqnarray}
\sigma^{NLO}_{\{n+1\}} & = & \int\limits_{n+1} \left( \left. d\sigma^R \right|_{\varepsilon=0}
- \left. d\sigma^A \right|_{\varepsilon=0} \right).
\end{eqnarray}
\item Analytic integrability (in $D$ dimensions) over the one-parton subspace leading to soft and collinear
divergences. This gives the contribution
\begin{eqnarray}
\sigma^{NLO}_{\{n\}} & = & \int\limits_n \left( d\sigma^V + \int\limits_1 d\sigma^A \right)_{\varepsilon=0}.
\end{eqnarray}
\end{itemize}
The final structure of an NLO calculation is then
\begin{eqnarray}
\label{NLOcrosssection}
\sigma^{NLO} & = & \sigma^B + \sigma^{NLO}_{\{n\}} + \sigma^{NLO}_{\{n+1\}} .
\end{eqnarray}
All contributions on the r.h.s of eq.(\ref{NLOcrosssection}) are now finite.
However, it should be noted that
$\sigma^{NLO}_{\{n+1\}}$ is a difference of two terms and therefore not guaranteed to be
positive-definite.
Similar, $\sigma^{NLO}_{\{n\}}$ contains the interference term between the one-loop amplitude
and the Born amplitude and is therefore also not guaranteed to be
positive-definite.  
\\
\\
The $(n+1)$ matrix element is approximated by a sum of dipole terms
\begin{eqnarray}
\lefteqn{\sum\limits_{pairs\; i,j} \sum\limits_{k \neq i,j} {\cal D}_{ij,k} =} \nonumber \\
& = & 
\sum\limits_{pairs\; i,j} \sum\limits_{k \neq i,j} 
- \frac{1}{2 p_i \cdot p_j}
\langle 1, ..., \tilde{(ij)},...,\tilde{k},...|
\frac{{\bf T}_k \cdot {\bf T}_{ij}}{{\bf T}^2_{ij}} V_{ij,k} |
1,...,\tilde{(ij)},...,\tilde{k},... \rangle, \nonumber \\
\end{eqnarray}
where the emitter parton is denoted by $\tilde{ij}$ and the spectator
by $\tilde{k}$.
Here ${\bf T}_i$ denotes the colour charge operator for parton $i$.
The colour charge operators ${\bf T}_i$ for a quark, gluon and antiquark in the final state are
\bq
\mbox{quark :} & & \l ... q_i ... | T_{ij}^a | ... q_j ... \r, \nonumber \\
\mbox{gluon :} & & \l ... g^c ... | i f^{cab} | ... g^b ... \r, \nonumber \\
\mbox{antiquark :} & & \l ... \bar{q}_i ... | \left( - T_{ji}^a \right) | ... \bar{q}_j ... \r.
\eq
$V_{ij,k}$ is a matrix in the helicity space of the emitter with the correct
soft and collinear behaviour.
$|1,...,\tilde{(ij)},...,\tilde{k},... \rangle$ is a vector in colour- and helicity space.
It is also convenient to introduce the variable
\bq
y_{ij,k} & = & \frac{2p_ip_j}{2p_ip_j+2p_ip_k+2p_jp_k}
\eq
and the ordered set ${\cal S}$ consisting of all relevant $y_{ij,k}$-variables:
\bq
{\cal S} & = & \left\{ y_{ij,k} | i \;\mbox{emitter}; j \;\mbox{emitted particle}; k \;\mbox{spectator} \right\}.
\eq
We denote the number of elements of ${\cal S}$ by $m$.
Explicit formulae for the dipole terms ${\cal D}_{ij,k}$ have been given by Catani and Seymour \cite{Catani:1997vz}. 
In the CDR-scheme they read \footnote{Corresponding formulae also exist for the four-dimensional
scheme as well as conversion rules between the two schemes \cite{Catani:1997pk}.}
\begin{eqnarray}
V_{q_i g_j,k} & = & 8 \pi \mu^{2\varepsilon} \alpha_s \delta_{s s'} C_F
\left[ \frac{2}{1-\tilde{z}_i (1-y_{ij,k})} - (1 + \tilde{z}_i) - \varepsilon
(1 -\tilde{z}_i ) \right] ,\nonumber \\
V_{q_i \bar{q}_j,k} & = & 8 \pi \mu^{2 \varepsilon} \alpha_s  T_R
\left[ - g^{\mu\nu} - \frac{2}{p_i p_j}
(\tilde{z}_i p_i^\mu - \tilde{z}_j p_j^\mu) ( \tilde{z}_i p_i^\nu - \tilde{z}_j p_j^\nu )
\right] ,\nonumber \\
V_{g_i g_j,k} & = & 8 \pi \mu^{2 \varepsilon} \alpha_s 2 C_A
\left[
-g^{\mu\nu} \left( \frac{1}{1-\tilde{z}_i(1-y_{ij,k})}
+ \frac{1}{1-\tilde{z}_j(1-y_{ij,k})} - 2 \right) \right. \nonumber \\
& & \left. + ( 1 - \varepsilon) \frac{1}{p_i p_j}
(\tilde{z}_i p_i^\mu - \tilde{z}_j p_j^\mu) ( \tilde{z}_i p_i^\nu - \tilde{z}_j p_j^\nu )
\right].
\end{eqnarray}
where $s,s'$ are the spin indices of the fermion $\tilde{ij}$ and $\mu,\nu$
are the spin indices of the gluon $\tilde{ij}$.
The momenta of the emitter and the spectator are related to the original $(n+1)$-parton
configuration by
\begin{eqnarray}
\tilde{p}_k^\mu & = & \frac{1}{1-y_{ij,k}} p_k^\mu , \nonumber \\
\tilde{p}_{ij}^\mu & =& p_i^\mu + p_j^\mu - \frac{y_{ij,k}}{1-y_{ij,k}} p_k^\mu.
\end{eqnarray}
The variables $\tilde{z}_i$ and $\tilde{z}_j$ are given by
\begin{eqnarray}
\tilde{z}_i = \frac{p_i p_k}{p_j p_k + p_i p_k}, & &
\tilde{z}_j = 1 - \tilde{z}_i.
\end{eqnarray}
We subtract the approximation only for $y_{ij,k} < y_{res}$.
Therefore
\begin{eqnarray}
\label{re1}
\lefteqn{
\int d \sigma^R - d \sigma^A } \nonumber \\
& = & \int d\phi_{n+1} \left|{\cal M}(p_1,...,p_{n+1})\right|^2 
  \Theta(\mbox{all}\; y_{ij,k}>y_{res})
  \Theta^{cut}_{n+1}(p_1,...,p_{n+1}) \nonumber \\
& & \mbox{} + \int d\phi_{n+1} 
 \left[ \left|{\cal M}(p_1,...,p_{n+1})\right|^2 
   \left( 1- \Theta(\mbox{all}\; y_{ij,k}>y_{res}) \right)
   \Theta^{cut}_{n+1}(p_1,...,p_{n+1}) \right. \nonumber \\
& & \left. - 
\sum\limits_{pairs\; i,j} \sum\limits_{k \neq i,j} {\cal D}_{ij,k}(p_1,...,p_{n+1}) 
 \Theta(y_{ij,k} < y_{res})
 \Theta^{cut}_{n}(p_1,...,\tilde{p}_{ij},...,\tilde{p}_k,...,p_{n+1}) \right] . \nonumber \\
\end{eqnarray}
Here we split the contribution from $d\sigma^R$ into a ``resolved'' region (first line)
and an ``unresolved'' region (second line).
The first line is infrared finite and positive definite.
The sum of the second and third line is by construction infrared finite, but not
necessarily positive definite.
Both $d\sigma^R$ and $d\sigma^A$ are integrated over the same $(n+1)$ parton phase space, but it should be noted that
$d\sigma^R$ is proportional to $\Theta^{cut}_{n+1}$, whereas $d\sigma^A$ is proportional to $\Theta^{cut}_n$.
Here $\Theta^{cut}_n$ denotes the jet-defining function for $n$-partons.
\\
\\
The subtraction term can be integrated over the one-parton phase space to yield the term
\bq
{\bf I} \otimes d\sigma^B & = & \int\limits_{1} d\sigma^A \Theta(y_{ij,k} < y_{res}) = \sum\limits_{pairs\; i,j} \sum\limits_{k \neq i,j} \int d\phi_{dipole} {\cal D}_{ij,k} \Theta(y_{ij,k} < y_{res}).
\eq
The universal factor ${\bf I}$ still contains colour correlations, but does not depend on the unresolved parton $j$.
The explicit results corresponding to the restricted region of integration $y_{ij,k}<y_{res}$
have been given by Nagy and Tr\'ocs\'anyi \cite{Nagy:1998bb}:
\bq
\lefteqn{
\int d\phi_{dipole} {\cal D}_{ij,k} \Theta(y_{ij,k} < y_{res}) } \nonumber \\
& = & 
 - \frac{\alpha_s}{2\pi} \frac{1}{\Gamma(1-\eps)} \left( \frac{4\pi \mu^2}{2\tilde{p}_{ij} \tilde{p}_k} \right)^\eps {\cal V}_{ij}(\eps,y_{res})
 \left\l {\cal M}_{Born}^{(n)} \left|
   \frac{{\bf T}_{ij} \cdot {\bf T}_k}{{\bf T}_{(ij)k}^2} \right| {\cal M}_{Born}^{(n)} \right\r
\eq
with
\bq
{\cal V}_{qg}(\eps,y_{res}) & = & 
 C_F \left\{ \left( \frac{1}{\eps^2} - \ln^2 y_{res} \right) 
             + \frac{3}{2} \left( \frac{1}{\eps} - 1 + y_{res} - \ln y_{res} \right)
             + 5 - \frac{\pi^2}{2} + O(\eps) \right\}, \nonumber \\ 
{\cal V}_{gg}(\eps,y_{res}) & = & 
 2 C_A \left\{ \left( \frac{1}{\eps^2} - \ln^2 y_{res} \right) 
               + \frac{11}{6} \left( \frac{1}{\eps} - 1 + y_{res} - \ln y_{res} \right)
               + \frac{50}{9} - \frac{\pi^2}{2} + O(\eps) \right\}, \nonumber \\
{\cal V}_{q\bar{q}}(\eps,y_{res}) & = & 
 T_R \left\{ - \frac{2}{3} \left( \frac{1}{\eps} - 1 + y_{res} - \ln y_{res} \right)
             - \frac{16}{9} + O(\eps) \right\}.
\eq
The term ${\bf I} \otimes d\sigma^B$ lives on the phase space of the $n$-parton configuration and has the appropriate
singularity structure to cancel the infrared divergences coming from the one-loop amplitude.
Therefore
\bq
d\sigma^V + {\bf I} \otimes d\sigma^B
\eq
is infrared finite.\\
\\
We now introduce a function $\Theta_{ab,c}$
with $\Theta_{ab,c} = 1$ if $y_{ab,c}$ is the smallest element
in the set ${\cal S}$, and $\Theta_{ab,c} = 0$ otherwise. 
If two invariants are equal (which is for example the case for $y_{ab,c}$ and $y_{ba,c}$)
we consider the invariant $y_{ab,c}$ with the smaller value of 
\bq
\frac{2p_bp_c}{2p_ap_b + 2p_ap_c + 2 p_b p_c}
\eq
to be smaller.
Since one of the elements must be the smallest, we have 
\bq
1 & = & \sum\limits_{{\cal S}} \Theta_{ab,c}.
\eq
The sum is
over all elements in the set ${\cal S}$. 
We further define a parameter $s_{min} = y_{min} Q^2$ and slice the region where
$y_{as,b}$ is the smallest element in the set into two parts as follows:
The first region is defined as the region
where $s_{as} > s_{min}$ and $s_{sb} > s_{min}$.
The second region is the complement of the first: 
$s_{as} < s_{min}$ or $s_{sb} < s_{min}$.
We now rewrite eq.(\ref{re1}) as follows:
\bq
\label{A123}
\int d \sigma^R - d \sigma^A = 
 \int d\phi_{n+1} 
 \sum\limits_{{\cal S}} \Theta_{as,b} 
 \left( A_1 + A_2 + A_3 \right)
\eq
with
\bq
\label{B123}
A_1 & = & \left|{\cal M}(p_1,...,p_{n+1})\right|^2 
  \Theta(\mbox{all}\; y_{ij,k}>y_{res})
  \Theta^{cut}_{n+1}(p_1,...,p_{n+1}), \nonumber \\
A_2 & = & \Theta(s_{as}-s_{min}) \Theta(s_{sb}-s_{min}) \nonumber \\
 & & \cdot
   \left[ \left|{\cal M}(p_1,...,p_{n+1})\right|^2 
     \left( 1- \Theta(\mbox{all}\; y_{ij,k}>y_{res}) \right)
     \Theta^{cut}_{n+1}(p_1,...,p_{n+1}) 
 \right. \nonumber \\ & & \left.
   - \sum\limits_{pairs\; i,j} \sum\limits_{k \neq i,j} {\cal D}_{ij,k}(p_1,...,p_{n+1}) 
     \Theta(y_{ij,k} < y_{res})
     \Theta^{cut}_{n}(p_1,...,\tilde{p}_{ij},...,\tilde{p}_k,...,p_{n+1}) \right], \nonumber \\
A_3 & = & \left[ 1- \Theta(s_{as}-s_{min}) \Theta(s_{sb}-s_{min}) \right] \nonumber \\
 & & \cdot
   \left[ \left|{\cal M}(p_1,...,p_{n+1})\right|^2 
     \left( 1- \Theta(\mbox{all}\; y_{ij,k}>y_{res}) \right)
     \Theta^{cut}_{n+1}(p_1,...,p_{n+1}) 
 \right. \nonumber \\ & & \left.
   - \sum\limits_{pairs\; i,j} \sum\limits_{k \neq i,j} {\cal D}_{ij,k}(p_1,...,p_{n+1}) 
     \Theta(y_{ij,k} < y_{res})
     \Theta^{cut}_{n}(p_1,...,\tilde{p}_{ij},...,\tilde{p}_k,...,p_{n+1}) \right]. \nonumber \\
\eq
Choosing $y_{min}$ small enough makes the contribution from the $A_3$-term negligible.
We will therefore not consider this term further.
From eq. (\ref{A123}) and eq. (\ref{B123}) the structure of our algorithm emerges.
If all $y_{ij,k}$ are larger than $y_{res}$, we are in the region corresponding to the $A_1$-term
and we generate configurations according to the unsubtracted matrix element
$\left| {\cal M}_{Born}^{(n+1)} \right|^2$.
If one $y_{ij,k}$ is smaller than $y_{res}$, we evaluate the integral of the unresolved region
corresponding to the $A_2$-term, combine it with the virtual corrections and generate a 
$n$-parton configuration according to this result.

\section{Review of Monte Carlo tools}

In this section we review some algorithms concerning the generation of phase space
as well as the Metropolis algorithm for sampling a given distribution.

\subsection{The Metropolis algorithm}

In practical applications one often wants to generate random variables
according to some probability density $P(x_1,...,x_d)$, which not
necesarrily factorizes. 
In practice one often uses the Metropolis algorithm 
(\cite{Metropolis}, \cite{Bhanot:1988sf}) for this purpose.
Let us call the vector $\phi=(x_1,...,x_d)$ a state of the ensemble, which
we want to generate.
Within the Metropolis algorithm one starts from a state $\phi_0$, and
replaces iteratively an old state by a new one, in such a way,
that the correct probability density distribution is obtained in the limit
of a large number of such iterations.
The equilibrium distribution is reached, regardless of the state one started
with. Once the equilibrium distribution is reached, repeated application
of the algorithm keeps one in the same ensemble.
In short, the desired distribution is the unique fix point of the algorithm.
Two important conditions have to be met for the Metropolis algorithm to work:
Ergodicity and detailed balance.
Detailed balance states that the transition probabilities 
$W(\phi_1 \rightarrow \phi_2)$ and $W(\phi_2 \rightarrow \phi_1)$
obey
\bq
P(\phi_1) W(\phi_1 \rightarrow \phi_2) & = & P(\phi_2) W(\phi_2 \rightarrow \phi_1).
\eq
Ergodicity requires that each state can be reached from any other state
within a finite number of steps. 
Given a state $\phi_1$, one iteration of the Metropolis algoritm
consists of the following steps:
\begin{enumerate}
\item Generate (randomly) a new candidate $\phi'$.
\item Calculate $\Delta S = - \ln(P(\phi')/P(\phi_1))$.
\item If $\Delta S < 0$ set the new state $\phi_2=\phi'$.
\item If $\Delta S > 0$ accept the new candidate only with
probability $P(\phi')/P(\phi)$, otherwise retain the old state
$\phi_2 = \phi_1$.
\item Do the next iteration.
\end{enumerate}
Step 3 and 4 can be summarized that the probability of accepting the 
candidate $\phi'$ is given by 
$W(\phi_1 \rightarrow \phi')=\mbox{min}(1,e^{-\Delta S})$.
It can be verified that this transition probabilty satisfies detailed
balance.
The way how a new candidate $\phi'$ suggested is arbitrary, 
restricted only by the condition that one has to be able to reach each
state within a finite number of steps as well as the condition that the
probability $A( \phi_1 \rightarrow \phi' )$ 
(the probability of suggesting $\phi'$ given we are in state $\phi_1$ )
has to be equal to the probability $A(\phi' \rightarrow \phi_1)$.

\subsection{Rambo}

The RAMBO-algorithm \cite{Kleiss:1986gy} maps a hypercube $[0,1]^{4n}$ of random numbers into $n$ 
physical four-momenta with center-of-mass energy $\sqrt{P^2}$.
Massless four-vectors can be generated with uniform weight.\\
\\
Let $P=(P,0,0,0)$ be a time-like four-vector. 
The phase space volume for a system of $n$ massless particles with
center-of-mass energy $\sqrt{P^2}$ is
\begin{eqnarray}
\label{phase_space_measure}
\Phi_n & = & \int \prod\limits_{i=1}^n \frac{d^4 p_i}{(2 \pi)^3} \theta(p_i^0) \delta(p_i^2)  
(2 \pi)^4 \delta^4\left( P - \sum\limits_{i=1}^n p_i \right).
\end{eqnarray}
To derive the RAMBO-algorithm one starts instead from the quantity
\begin{eqnarray}
R_n & = & \int \prod\limits_{i=1}^n \frac{d^4 q_i}{(2 \pi)^3} \theta(q_i^0) \delta(q_i^2)  
(2 \pi)^4 f(q_i^0) 
= (2 \pi)^{4-2n} \left( \int\limits_0^\infty x f(x) dx \right)^n.
\end{eqnarray}
The quantity $R_n$ can be interpreted as describing a system of $n$ massless four-momenta $q_i^\mu$
that are not constrained by momentum conservation but occur with some weight function $f$ which
keeps the total volume finite.
The four-vectors $q_i^\mu$ are then related to the physical four-momenta $p_i^\mu$ by the
following Lorentz and scaling transformations:
\begin{eqnarray}
\label{boost_and_scale}
p_i^0 = x \left( \gamma q_i^0 + \vec{b} \cdot \vec{q}_i \right), & &
\vec{p}_i = x \left( \vec{q}_i + \vec{b} q_i^0 + a \left( \vec{b} \cdot \vec{q}_i \right) \vec{b} \right),
\end{eqnarray}
where
\begin{eqnarray}
& & Q^\mu = \sum\limits_{i=1}^n q_i^\mu, \;\;\; M = \sqrt{Q^2}, \;\;\; 
\vec{b} = - \frac{1}{M} \vec{Q}, \nonumber \\
& & \gamma = \frac{Q^0}{M} = \sqrt{1 + \vec{b}^2}, \;\;\; a = \frac{1}{1+\gamma}, \;\;\; x = \frac{\sqrt{P^2}}{M}.
\end{eqnarray}
Denote this transformation and its inverse as follows
\begin{eqnarray}
p_i^\mu = x H^\mu_{\vec{b}}(q_i), & & q_i^\mu = \frac{1}{x} H^\mu_{-\vec{b}}(p_i).
\end{eqnarray}
By a change of variables one can show
\begin{eqnarray}
R_n & = & \int \prod\limits_{i=1}^n \left( \frac{d^4 p_i}{(2 \pi)^3} \delta(p_i^2) \theta(p_i^0) \right)
(2 \pi)^4 \delta^4 \left(P - \sum\limits_{i=1}^n p_i \right) \nonumber \\
& &\cdot \left( \prod\limits_{i=1}^n f \left( \frac{1}{x} H^0_{-\vec{b}}(p_i) \right) \right)
\frac{(P^2)^2}{x^{2n+1} \gamma} d^3b dx.
\end{eqnarray}
With the choice $f(x) = e^{-x}$ the integrals over $\vec{b}$ and $x$ may be performed and 
one obtains
\begin{eqnarray}
R_n & = & \Phi_n \cdot S_n
\end{eqnarray}
with 
\begin{eqnarray}
S_n & = & 2 \pi (P^2)^{2-n} \frac{ \Gamma\left(\frac{3}{2}\right) \Gamma(n-1) \Gamma(2n)}{\Gamma\left(n+\frac{1}{2}\right)}.
\end{eqnarray}
This gives a Monte Carlo algorithm which generates massless four-momenta $p_i^\mu$ according to the phase-space
measure (\ref{phase_space_measure}).
The algorithm consists of two steps:
\begin{enumerate}
\item Generate independently $n$ massless four-momenta $q_i^\mu$ with isotropic angular distribution and 
energies $q_i^0$ distributed according to the density $q_i^0 e^{-q_i} dq_i^0$.
Using $4n$ random numbers $u_i$ uniformly distributed in $\left[0,1\right]$ this is done as follows:
\begin{eqnarray}
c_i = 2 u_{i_1} -1, & \varphi_i = 2 \pi u_{i_2}, & q_i^0 = - \ln(u_{i_3} u_{i_4}), \nonumber \\
q_i^x = q_i^0 \sqrt{1- c_i^2} \cos \varphi_i, & q_i^y = q_i^0 \sqrt{1-c_i^2} \sin \varphi_i, & q_i^z = q_i^0 c_i.
\end{eqnarray}
\item The four-vectors $q_i^\mu$ are then transformed into the four-vectors $p_i^\mu$, using the transformation
(\ref{boost_and_scale}).
\end{enumerate}
Each event has the uniform weight
\bq
\label{masslessweight}
w_0 & = & (2 \pi)^{4-3n} \left( \frac{\pi}{2} \right)^{n-1} \frac{(P^2)^{n-2}}{\Gamma(n) \Gamma(n-1)}.
\eq
In summary, the RAMBO algorithm provides a mapping from a $4n$-dimensional
hypercube into $n$ physical four-momenta, each event being generated with weight $w$.
By picking points in the hypercube at random, we sweep out the complete phase space.
Therefore ergodicity is satisfied and we can combine the RAMBO algorithm with the
Metropolis algorithm.

\subsection{Generating configurations close to soft or collinear regions}

In this section we review an algorithm on how to obtain a $(n+1)$-parton configuration
from a given $n$-parton configuration.
An additional four-momentum $p_s$ is inserted between the legs $p_a$ and $p_b$. 
For efficiency we generate the events such that $s_{as}, s_{sb} > s_{min}$ with
$s_{min} = y_{min} Q^2$.
We assume that the complementary region $s_{as} < s_{min}$ or $s_{sb} < s_{min}$
gives only a contribution of order $y_{min}$.
We first relate a given $(n+1)$-configuration to a $n$-parton configuration and invert
then this relation to obtain an algorithm to generate soft or collinear insertions
\cite{Weinzierl:1999yf}.\\
\\
Let $k_a'$, $k_s$ and $k_b'$ be the momenta of the $(n+1)$-parton configuration such that $s_{as} = (k_a' + k_s)^2$, 
$s_{sb} = (k_b' + k_s)^2$ and $s_{ab} = (k_a' + k_s + k_b')^2$. We want to relate this $(n+1)$ particle
configuration to a nearby ``hard'' $n$-particle configuration with $(k_a + k_b)^2 = (k_a' + k_s + k_b')^2$,
where $k_a$ and $k_b$ are the corresponding ``hard'' momenta.
Using the factorization of the phase space, we have 
\begin{eqnarray}
d\Phi_{n+1} & = & d\Phi_{n-1} \frac{dK^2}{2 \pi} d\Phi_3(K,k_a',k_s,k_b').
\end{eqnarray}
The three-particle phase space is given
by
\begin{eqnarray}
d\Phi_3(K,k_a',k_s,k_b') & = & \frac{1}{32 (2 \pi)^5 s_{ab}} ds_{as} ds_{sb} d\Omega_b' d\phi_s \nonumber \\
& = & \frac{1}{4 (2 \pi)^3 s_{ab}} ds_{as} ds_{sb} d\phi_s d\Phi_2(K,k_a,k_b)
\end{eqnarray}
and therefore
\begin{eqnarray}
d\Phi_{n+1} & = & d\Phi_n \frac{ds_{as} ds_{sb} d\phi_s}{4 (2 \pi)^3 s_{ab} }.
\end{eqnarray}
The region of integration for $s_{as}$ and $s_{sb}$ is 
$s_{as} > s_{min}$, $s_{sb} > s_{min}$ (since we want to restrict the integration to the region
where the invariants are larger than $s_{min}$)
and $s_{as} + s_{sb} < s_{ab}$ (Dalitz plot for massless particles).
It is desirable to absorb poles in $s_{as}$ and $s_{sb}$ into the measure. 
A naive numerical integration of these poles without any remapping
results in a poor accuracy.
This is done
by changing the variables according to
\begin{eqnarray}
s_{as} = s_{ab} \left( \frac{s_{min}}{s_{ab}} \right)^{u_1}, & & 
s_{sb} = s_{ab} \left( \frac{s_{min}}{s_{ab}} \right)^{u_2},
\end{eqnarray}
where $0 \leq u_1, u_2 \leq 1$. Note that $u_1, u_2 > 0$ enforces $s_{as}, s_{sb} > s_{min}$.
Therefore this transformation of variables may only be applied to invariants $s_{ij}$ 
where the region $0 < s_{ij} < s_{min}$ is cut out.
The phase space measure becomes
\begin{eqnarray}
d\Phi_{n+1} & = & d\Phi_{n} \frac{1}{4 (2 \pi)^3} \frac{s_{as} s_{sb}}{s_{ab}} \ln^2\left(\frac{s_{min}}{s_{ab}}\right)
\Theta(s_{as} + s_{sb} < s_{ab} ) du_1 du_2 d\phi_s .
\end{eqnarray}
This give the following algorithm for generating a $(n+1)$-parton configuration:
\begin{enumerate}
\item Take a ``hard'' $n$-parton configuration and pick out two momenta $k_a$ and $k_b$. Use three
uniformly distributed random number $v_1,v_2,v_3$ and set
\begin{eqnarray}
s_{ab} & = & (k_a + k_b)^2 ,\nonumber \\
s_{as} & = & s_{ab} \left( \frac{s_{min}}{s_{ab}} \right)^{v_1} ,\nonumber \\
s_{sb} & = & s_{ab} \left( \frac{s_{min}}{s_{ab}} \right)^{v_2} ,\nonumber \\
\phi_s & = & 2 \pi v_3.
\end{eqnarray}
\item If $(s_{as} + s_{sb} ) > s_{ab} $, reject the event.
\item If not, solve for $k_a'$, $k_b'$ and $k_s$.
If $s_{as} < s_{sb}$ we want to have $k_b' \rightarrow k_b$ as $s_{as} \rightarrow 0$.
Define
\begin{eqnarray}
E_a = \frac{s_{ab} - s_{sb}}{2 \sqrt{s_{ab}}}, \;\;\;
E_s = \frac{s_{as}+s_{sb}}{2 \sqrt{s_{ab}}} ,\;\;\;
E_b = \frac{s_{ab} - s_{as}}{2 \sqrt{s_{ab}}}, 
\end{eqnarray}
\begin{eqnarray}
\theta_{ab}  =  \arccos \left( 1 -\frac{s_{ab} - s_{as} - s_{sb}}{2 E_a E_b} \right), & &
\theta_{sb}  =  \arccos \left( 1 - \frac{s_{sb}}{2 E_s E_b} \right) .
\end{eqnarray}
It is convenient to work in a coordinate system which is obtained by a Lorentz transformation to the
center of mass of $k_a+k_b$ and a rotation such that $k_b'$ is along the positive $z$-axis. In that
coordinate system
\begin{eqnarray}
p_a' & = & E_a ( 1, \sin \theta_{ab} \cos(\phi_s+\pi), \sin \theta_{ab} \sin(\phi_s+\pi), \cos \theta_{ab} ) ,\nonumber \\
p_s & = & E_s ( 1, \sin \theta_{sb} \cos \phi_s, \sin \theta_{sb} \sin \phi_s, \cos \theta_{sb} ) ,\nonumber \\
p_b' & = & E_b ( 1, 0, 0, 1) .
\end{eqnarray}
The momenta $p_a'$, $p_s$ and $p_b'$ are related to the momenta $k_a'$, $k_s$ and $k_b'$ by a sequence of
Lorentz transformations back to the original frame
\begin{eqnarray}
k_a' & = & \Lambda_{boost} \Lambda_{xy}(\phi) \Lambda_{xz}(\theta) p_a'
\end{eqnarray}
and analogous for the other two momenta. 
The explicit formulae for the Lorentz transformations are obtained as follows :\\
\\
Denote by $K = \sqrt{(k_a+k_b)^2}$ and by $p_b$ the coordinates of the hard momentum $k_b$ in the center of
mass system of $k_a+k_b$. $p_b$ is given by
\begin{eqnarray}
p_b & = & \left( 
\frac{Q^0}{K} k_b^0 - \frac{\vec{k}_b \cdot \vec{Q}}{K}, \vec{k}_b + \left( \frac{\vec{k}_b \cdot \vec{Q}}{K (Q^0+K)} 
  - \frac{k_b^0}{K} \right) \vec{Q}
\right)
\end{eqnarray}
with $Q = k_a + k_b$. The angles are then given by
\begin{eqnarray}
\theta & = & \arccos \left( 1 - \frac{p_b \cdot p_b'}{2 p_b^t p_b^{t'}} \right), \nonumber \\
\phi & = & \arctan\left( \frac{p_b^y}{p_b^x} \right).
\end{eqnarray}
The explicit form of the rotations is
\begin{eqnarray}
\Lambda_{xz}(\theta) & = & \left(
\begin{array}{cccc}
1 & 0 & 0 & 0 \\
0 & \cos \theta & 0 & \sin \theta \\
0 & 0 & 1 & 0 \\
0 & - \sin \theta & 0 & \cos \theta \\
\end{array}
\right), \nonumber \\
\Lambda_{xy} (\phi) & = & 
\left(
\begin{array}{cccc}
1 & 0 & 0 & 0 \\
0 & \cos \phi & - \sin \phi & 0 \\
0 & \sin \phi & \cos \phi & 0 \\
0 & 0 & 0 & 1 \\
\end{array}
\right).
\end{eqnarray}
The boost $k' = \Lambda_{boost} q $ is given by 
\begin{eqnarray}
k' & = & \left( 
\frac{Q^0}{K} q^0 + \frac{\vec{q} \cdot \vec{Q}}{K}, \vec{q}+ \left( \frac{\vec{q} \cdot \vec{Q}}{K (Q^0+K)} 
  + \frac{q^0}{K} \right) \vec{Q}
\right)
\end{eqnarray}
with $Q = k_a + k_b$ and $K = \sqrt{(k_a+k_b)^2}$.
\item If $s_{as} > s_{sb}$, exchange $a$ and $b$ in the formulae above.
\item The ``soft'' event has then the weight
\begin{eqnarray}
w_{n+1} & = & \frac{\pi}{2} \frac{1}{(2 \pi)^3} \frac{s_{as} s_{sb}}{s_{ab}} \ln^2 \left( \frac{s_{min}}{s_{ab}} \right) w_n,
\end{eqnarray}
where $w_n$ is the weight of the original ``hard'' event.
\end{enumerate}
Again by picking $v_1$, $v_2$ and $v_3$ randomly, we sample the complete phase space
except the region $s_{as} < s_{min}$ or $s_{sb}< s_{min}$.

\section{Generating unweighted events according to leading-order matrix elements}

In this section we give an algorithm how to generate a sequence of unweighted
events according to leading-order matrix elements.
This section serves as a warm-up exercise to the next section, where we extend
the algorithm to next-to-leading order matrix elements.
Since for leading-order calcultions the amplitude squared is always positive,
negative weights do not occur at this stage.
However we would like to keep the discussion quite general and process-independent.
We therefore do not make any assumptions on the form of the matrix element squared.
In particular we do not assume that the amplitude squared, viewed as a multi-dimensional
probability density factorizes nor do we assume that we have at our disposal  
a simple function which is everywhere larger than the amplitude squared.
Therefore the acceptance-rejection method is not at our disposal and we use the
Metropolis algorithm to generate the desired distribution.\\
\\
Observables in high-energy collider experiments are often of the following
form
\bq
\label{obsLO}
\l O \r & = & \int d\Phi_n(p_a+p_b,p_1,...,p_n) \; \frac{\left|{\cal M}\right|^2 }{8 K(s)} \; J(O,p_1,...,p_n),
\eq
where $p_a$ and $p_b$ are the momenta of the incoming particles, the outgoing
particles are denoted by the labels $1$,...,$n$. The Lorentz-invariant phase space
is denoted by $d\Phi_n$, $1/(8K(s))$ is a kinematical factor with $s=(p_a+p_b)^2$ which
includes the averaging over initial spins (we assumed two spin states for each initial
particle), $\left| {\cal M} \right|^2$ is the relevant matrix element squared and $J(O,p_1,...,p_n)$
is a function which defines the observable and includes all experimental cuts.
The situation described above corresponds to electron-positron annihilation. If
hadrons appear in the initial state there are slight modifications.
Using the RAMBO algorithm eq. (\ref{obsLO}) may be rewritten as
\bq
\l O \r & = & \int d^{4n}u  \; w_0(u) \frac{\left|{\cal M}(p_i(u))\right|^2}{8 K(s)} \; J(O,p_1,...,p_n),
\eq
where the weight $w_0(u)$ is given by eq.(\ref{masslessweight}) and the final-state four-momenta
$p_i$ are obtained from the $u_j$'s through the RAMBO algorithm.
Suppose now that we have at our disposal a jet-defining function $\Theta_{y_{res}}^n(p_1,...,p_n)$ such that the support for $J(O,p_1,...,p_n)$ is contained in the support
for $\Theta_{y_{res}}^n(p_1,...,p_n)$. 
The function $\Theta_{y_{res}}^n(p_1,...,p_n)$ returns 1 if the $n$ partons are resolved at the scale
$y_{res}$ and 0 otherwise.
We may then replace $J(O,p_1,...,p_n)$
by
\bq
\Theta_{y_{res}}^n(p_1,...,p_n) J(O,p_1,...,p_n)
\eq
and generate random numbers $u_j$ distributed according to the 
probability density function
\bq
\label{LOprobdens}
\frac{1}{\sigma_0} \frac{1}{8 K(s)} w_0(u) \left|{\cal M}(p_i(u)\right|^2 \Theta_{y_{res}}^n(p_1,...,p_n),
\eq
where $\sigma_0$ is the total LO-cross section, which normalizes the probability density function
to unity:
\bq
\sigma_0 & = & \frac{1}{8 K(s)} \int d^{4n}u  \; w_0(u) \left|{\cal M}(p_i(u)\right|^2 \Theta_{y_{res}}^n(p_1,...,p_n).
\eq
Note that the quantity in eq. (\ref{LOprobdens}) is positive-definite, as required for a probability
density function.
We can apply the Metropolis algorithm to generate the desired distribution \footnote{The Metropolis algorithm has also been considered by Kharraziha and S. Moretti \cite{Kharraziha:1999iw} for the generation of phase space.
These authors use a different method to suggest a new candidate and do not combine the Metropolis
algorithm with a ``standard'' algorithm to generate the phase space.}.
\begin{enumerate}
\item To suggest a new candidate we randomly pick a point $(u_1',...,u_{4n}')$
in the $4n$-dimen\-sional hypercube and
use the RAMBO algorithm to obtain the corresponding $n$-parton configuration of four-vectors $p_1'$, ..., $p_n'$.
\item The probability density $P(u')$ for this candidate is given by eq. (\ref{LOprobdens}).
\item If $P(u')=0$ (which can occur due the presence of the jet-defining function\\
$\Theta_{y_{res}}^n(p_1',...,p_n')$) we reject the candidate and go back to step 1.
\item Calculate $\Delta S = -\ln(P(u')/P(u))$, where $P(u)$ is the probability density of the previous state.
\item If $\Delta S < 0$ accept the new candidate and return the $n$-parton configuration
$p_1'$, ..., $p_n'$.
\item If $\Delta S > 0$ accept the new candidate only with probability $P(u')/P(u)$,
otherwise retain the old state and return the old $n$-parton configuration $p_1$,...,$p_n$.
\item Do the next iteration.
\end{enumerate}
A technical comment on step 3 is in order: Step 3 immediately rejects candidates which do 
not pass the cuts $\Theta^n_{y_{res}}$.
This does not violate detailed balance due to the specific way new candidates are suggested
in step 1: A new state $(u_1',...,u_{4n}')$ is suggested independently of the present
state $(u_1,...,u_{4n})$.
If one uses a different procedure in step 1, for example by changing the $k$-th coordinate
$u_k \rightarrow u_k + \delta$ by a small, but random value $\delta$, candidates suggested
from states close to the bounday of $\Theta^n_{y_{res}}$ are more likely to fall
outside the boundary then candidates suggested from states inside the region.
To ensure detailed balance, one has to discard step 3 and treat candidates with zero 
probability in step 6 as candidates which are never accepted.
This inserts the old state once again into the event record.

\section{Generating unweighted events according to next-to-leading-order matrix elements}

In this section we generalize the algorithm from the previous section to next-to-leading
order.
For the real emission part we start from eq. (\ref{A123}) and eq. (\ref{B123}) and
first describe a method, how to pick two legs $a$ and $b$ corresponding to the emitter and the spectator.
We then try to insert a soft particle in between these two.
We divide the interval $[0,1]$ into $m$ subintervals ($m$ is the number of elements in the
set ${\cal S}$) and denote the size
of the $l$-th subinterval by $P_l$.
We further introduce functions $\Theta_l(x)$ which return $1$ if $x$ is in the $l$-th subinterval
and zero otherwise.
We may therefore rewrite eq. (\ref{A123}) as
\bq
\int d \sigma^R - d \sigma^A = 
 \int dx \; 
 \sum\limits_{l} \frac{1}{P_l} \Theta_l(x)
 \int d\phi_{n+1} 
  \Theta_{as,b} 
 \left( A_1 + A_2 \right).
\eq
Now the value of $x$ decides where the soft leg gets inserted.
In more detail, we may rearrange the above equation as
\bq
\int d \sigma^R - d \sigma^A = 
\int d^{4n}u\; w_0(u) \int dx\; 
 \sum\limits_{l} \frac{1}{P_l} \Theta_l(x)
\int dv_1 \; dv_2 \; dv_3 \; w_{soft} 
  \Theta_{as,b} 
 \left( A_1 + A_2 \right), \nonumber \\
\eq
with
\bq
w_{soft} & = & \frac{\pi}{2} \frac{1}{(2 \pi)^3} \frac{s_{as} s_{sb}}{s_{ab}} 
\ln^2 \left( \frac{s_{min}}{s_{ab}} \right).
\eq
We may therefore use $4n+1+3$ random numbers $u_1$,...,$u_{4n}$, $x$, $v_1$, $v_2$ and $v_3$,
all uniformly distributed in $[0,1]$ and generate the phase space as follows:
We use the random numbers $u_1$,...,$u_{4n}$ to generate a hard configuration of $n$
four-momenta using the RAMBO algorithm.
The value of the random number $x$ decides where we try to insert a soft leg.
We check into which of the $m$ subintervals $x$ falls and pick out the corresponding
element of the set ${\cal S}$. This tells us which parton is going to be the emitter
and which parton is going to be the spectator.
We now use the random numbers $v_1$, $v_2$ and $v_3$ and the algorithm for soft
insertions and generate a $(n+1)$-parton configuration, where a soft particle
has been inserted between the emitter and the spectator.
This suggests the following Metropolis algorithm:
\begin{enumerate}
\item To suggest a new candidate we randomly pick a point $(u_1',...,u_{4n}',x',v_1',v_2',v_3')$
in the $4n+1+3$-dimensional hypercube.
Using the first $4n$ random numbers $u_1$,...,$u_{4n}$ and 
the RAMBO algorithm we obtain the corresponding ``hard'' $n$-parton configuration of four-vectors $\phi_{hard}'=(q_1'$, ..., $q_n')$.
From this hard configuration and the values of $x$ and $v_1$, $v_2$ and $v_3$ we obtain
a related ``soft'' $(n+1)$-parton configuration
$\phi_{soft}'=(p_1'$, ..., $p_n',p_{n+1}')$ together with the information where
the soft particle $s$ has been inserted. Therefore the emitter $a$ and the spectator $b$
are known.
\item Check if $y_{as,b}$ is the smallest value in the set ${\cal S}$. If not, reject the candidate and go back to step 1.
\item Check if all $y_{ij,k}$ are larger then $y_{res}$.\\
\\
If this is the case, the probabilty density of the new candidate is given by
\bq
P' & = &
\frac{1}{\sigma_{NLO}} \frac{1}{8 K(s)} \frac{w_0(u) w_{soft}}{P_l} \left|{\cal M}_{Born}^{(n+1)}(p_1',...,p_{n+1}')\right|^2 \Theta_{y_{res}}^{n+1}(p_1',...,p_{n+1}').
\eq
and we set a flag $f=1$.
Here $\sigma_{NLO}$ is given by
\bq
\lefteqn{
\sigma_{NLO} = 
 \frac{1}{8K(s)} } \nonumber \\
& &  \cdot \left\{ \int d\phi_n
   \left[ \left| {\cal M}_{Born}^{(n)} \right|^2 
        + 2 \; \mbox{Re}\; {{\cal M}_{Born}^{(n)}}^\ast {\cal M}_{1-loop}^{(n)} 
        + \langle {\cal M}_{Born}^{(n)} | {\bf I } | {\cal M}_{Born}^{(n)} \rangle 
 \right] \Theta^n_{y_{res}} \right. \nonumber \\
& & \left. + \int d\phi_{n+1} \left[ \left| {\cal M}_{Born}^{(n+1)} \right|^2 \Theta^{n+1}_{y_{res}}
     - \sum\limits_{pairs\; i,j} \sum\limits_{k \neq i,j} {\cal D}_{ij,k} 
           \Theta(y_{ij,k} < y_{res}) \Theta^n_{y_{res}} \right] \right\}.
\eq
$\Theta^{n+1}_{y_{res}}$ returns 1 if at least $n$ partons are resolved 
at the scale $y_{res}$.\\
If $P'=0$ (which can occur due the presence of the jet-defining function\\
$\Theta_{y_{res}}^{n+1}(p_1',...,p_{n+1}')$) we reject the candidate and go back to step 1.\\
\\
If at least one $y_{ij,k}$ in the set ${\cal S}$ is smaller then $y_{res}$
the probability density 
is given by
\bq
\label{probnlo}
\lefteqn{
P' =  
\frac{1}{\sigma_{NLO}} \frac{1}{8 K(s)} \frac{w_0}{V} } \nonumber \\
& & \left\{
 \left[ \left| {\cal M}_{Born}^{(n)} \right|^2 
        + 2 \; \mbox{Re}\; {{\cal M}_{Born}^{(n)}}^\ast {\cal M}_{1-loop}^{(n)} 
        + \langle {\cal M}_{Born}^{(n)} | {\bf I } | {\cal M}_{Born}^{(n)} \rangle 
 \right] \Theta^n_{y_{res}}(q_1',...,q_n') + R \right\}, \nonumber \\
\eq
where the integral over the unresolved region is given by
\bq
R & = &
 \sum\limits_{\cal S} \int dv_1 \; dv_2 \; dv_3 \;
     w_{soft} \Theta_{as,b} 
     \left[ \left| {\cal M}_{Born}^{(n+1)} \right|^2 \left(1- \Theta(\mbox{all}\;y_{ij,k} > y_{res}) \right) 
            \Theta^{n+1}_{y_{res}}
 \right. \nonumber \\ & & \left.
     - \sum\limits_{pairs\; i,j} \sum\limits_{k \neq i,j} {\cal D}_{ij,k} 
           \Theta(y_{ij,k} < y_{res}) \Theta^n_{y_{res}}
     \right]
\eq
and the unresolved volume is given by
\bq
V & = & \sum\limits_{\cal S} \int dv_1 \; dv_2 \; dv_3 \; \Theta_{as,b}
\left(1- \Theta(\mbox{all}\;y_{ij,k} > y_{res}) \right).
\eq
In addition set the flag $f=0$.\\
If $P' < 0$ (which can occur if $y_{res}$ is chosen too small) throw an exception.
\item Calculate $\Delta S = -\ln(P'/P)$, where $P$ is the probability density of the previous state.
\item If $\Delta S < 0$ accept the new candidate.
If the flag $f=1$ return the ``soft'' $(n+1)$-parton configuration
$p_1'$, ..., $p_n'$ otherwise if $f=0$ return the ``hard'' $n$-parton configuration
$q_1',...,q_n'$.
\item If $\Delta S > 0$ accept the new candidate only with probability $P'/P$,
otherwise retain the old state and return the old parton configuration.
\item Do the next iteration.
\end{enumerate}
If the value of $y_{res}$ is chosen too small, it can occur that
$P'$ in eq. (\ref{probnlo})
turns out negative. However, this is not an artefact of our algorithm, but a failure
of perturbation theory: $P'$ corresponds to the NLO cross section for observing
$n$ jets with (fixed) four-vectors $q_1'$, ..., $q_n'$ resolved at a scale $y_{res}$.
If $y_{res}$ is too small the NLO prediction might become negative.
This is of course unphysical. The algorihtm throws an exception in this case
and one should rerun the program with a larger value of $y_{res}$.\\
\\
Given a generated sequence of momentum configurations $\phi_1$,...,$\phi_N$, where each
momentum configuration contains either $n$ or $(n+1)$ four-vectors, the Monte
Carlo estimate for an infrared safe observable is then given by
\bq
\l O \r & = & \sigma_{NLO} \frac{1}{N} \sum\limits_{i=1}^N J(O,\phi_i).
\eq
Here $J(O,\phi_i)$ is a function which defines the observable and includes all experimental
cuts.
The approximations involved are:
\begin{itemize}
\item If within an $(n+1)$-parton configuration some value of $y_{ij,k}$ is smaller
then $y_{res}$, this configuration is replaced with an $n$-parton configuration, which corresponds
to the average over the unresolved region combined with the virtual corrections and the Born term.
\item Due to the presence of the jet-defining functions $\Theta^n_{y_{res}}$ 
and $\Theta^{n+1}_{y_{res}}$ 
configurations which would classify as
$(n-1)$-parton events according to these functions are cut out.
One has to ensure that the support for the actual observable $O$ is contained in the support
for $\Theta^n_{y_{res}}$ 
and $\Theta^{n+1}_{y_{res}}$.
\item We neglected the term $A_3$ in eq. (\ref{A123}). This introduces a (negligible) systematic
error of $y_{min}$. This is done for efficiency.
\end{itemize}

\section{Conclusions}

In this paper we gave an algorithm to generate unweighted events of $n$ or $(n+1)$ four-vectors
according to next-to-leading order matrix elements.
We considered massless QCD in electron-positron annihilation. The extension of our methods to initial-state
partons or massive partons should be possible.

%\bibliography{/home/galileo/stefanw/notes/biblio}
%\bibliographystyle{/home/galileo/stefanw/latex/h-physrev3}

\end{document}